# Enhanced Room Temperature Coefficient of Resistance and Magneto-resistance of Ag-added $La_{0.7}Ca_{0.3-x}Ba_xMnO_3$ Composites

Rahul Tripathi[1, 3], V.P.S. Awana[1,*], Neeraj Panwar[2], G.L. Bhalla[3], H.U. Habermier[4], S.K.Agarwal[1] and H. Kishan[1],

[1]National Physical Laboratory, Dr. K. S. Krishnan Marg, New Delhi-110012, India.

[2]Department of Physics, Indian Institute of Technology, New Delhi-110016, India.

[3]Department of Physics and Astrophysics, University of Delhi, Delhi-1100007, India.

[4]Max Planck Institut fur Festkoerperforschung, Heisenbergstrasse-70569, Stuttgart, Germany.

**ABSTRACT**

In this paper we report an enhanced temperature coefficient of resistance (TCR) close to room temperature in $La_{0.7}Ca_{0.3-x}Ba_xMnO_3$ + $Ag_y$ (x = 0.10, 0.15 and $0 \leq y \leq 1.0$) (LCBMO+Ag) composite manganites. The observed enhancement of TCR is attributed to the grain growth and opening of new conducting channels in the composites. Ag addition has also been found to enhance intra-granular magneto-resistance. Inter-granular MR, however, is seen to decrease with Ag addition. The enhanced TCR and MR at / near room temperature open up the possibility of the use of such materials as infrared bolometric and magnetic field sensors respectively.

---

* Corresponding Author: Dr. V.P.S. Awana
National Physical Laboratory, Dr. K.S. Krishnan Marg, New Delhi-110012, India
Fax No. 0091-11-45609310: Phone no. 0091-11-45609210
e-mail-awana@mail.nplindia.ernet.in: www.freewebs.com/vpsawana/

## INTRODUCTION

Perovskite manganites with chemical formula $A_{1-x}B_xMnO_3$ (where A = rare earth element like La, Pr and B = Alkaline earth metals like Ca, Ba) are important both for the basic physics due to their rich magnetic and electronic phase diagrams, and for their practical applications aspects in magnetic field sensors and infra-red bolometers etc. [1-6]. So far the focus of most of the studies has been to observe enhanced magneto-resistance (MR) close to room temperature and in low magnetic field values. To achieve such characteristics, composites of these manganites containing insulating or metallic oxides have been synthesized [7-15]. Generally, within a particular range of doping, insulator-metal (I-M) transition at a temperature $T_P$ is observed near paramagnetic-ferromagnetic (PM-FM) transition temperature ($T_C$). The occurrence of the ferromagnetic metallic behaviour and the ensuing magneto-resistance below $T_C$ have been explained by the double-exchange (DE) model coupled with electron-phonon coupling [16]. The sharpness of the I-M transition can be useful in utilizing these manganite materials as infra-red / bolometer sensors. Goyal et al.[17] have shown that the thin films of various manganites have potential to be used for bolometer in a wide temperature range. Steeper is the I-M transition better is the temperature sensitivity of material. The sharpness of the electrical resistivity near I-M transition is described in terms of the temperature coefficient of resistance (TCR), defined as TCR % = [(1/R) (dR/dT)]x100, where R is the electrical resistance and T is the temperature. TCR is maximum near insulator-metal transition temperature $T_P$. Tuning of $T_P$ near room temperature is, therefore, important from the application point of view. However, it has been reported that as $T_P$ increases towards room temperature, TCR in manganites decreases [5, 17]. Therefore, one needs to optimize both $T_P$ and TCR near room temperature. In one of our previous studies TCR ~ 10% was achieved below 260K with Ag addition in polycrystalline $La_{0.67}Ca_{0.33}MnO_3$ [4].

To further observe $T_P$ close to room temperature with enhanced TCR, Ba was partially doped at Ca-site in $L_{0.7}Ca_{0.3}MnO_3$ and its composites were synthesized with Ag viz. $La_{0.7}Ca_{0.3-x}Ba_xMnO_3$+ $Ag_y$ [x = 0.10, 0.15; 0 ≤ y ≤ 1.0]. We report here the TCR and MR results of such composites.

## EXPERIMENTAL

Samples of $La_{0.7}Ca_{0.3-x}Ba_xMnO_3$ (0≤ x ≤ 0.3), $La_{0.7}Ca_{0.2}Ba_{0.1}MnO_3$+$Ag_y$ and $La_{0.7}Ca_{0.2}Ba_{0.1}MnO_3$+$Ag_y$ (0 ≤ y ≤ 1.0) composites were synthesized by the conventional solid-state reaction route using fine powders of $La_2O_3$, $CaCO_3$, $BaCO_3$, $MnO_2$ and Ag metal. The mixed powders were calcined at 1000°C, 1100°C and 1200°C in air for 24 hrs followed by thorough grinding each time. These powders were pre-sintered at 1300°C in air for 24 hrs. Finally the pelletized ceramics were annealed in air for 24 hrs at 1400°C. For oxygen intake, these pellets were annealed in flowing oxygen at 1100°C for 12 hrs and subsequently furnace cooled to room temperature. The structure and phase purity of the samples were analysed by powder x–ray diffraction (XRD) using Rigaku mini-flex diffractometer (λ = 1.54Å). The morphological studies have been carried out using LEO SEM 440 scanning electron microscope. ρ(T) measurements from 20K to 320K were carried out using the conventional four-probe method. MR measurements from room temperature down to 77K were carried out in an applied magnetic field of 0.3T.

## RESULTS AND DISCUSSION

Fig. 1 shows room temperature x-ray diffraction (XRD) patterns of the series $La_{0.7}Ca_{0.3-x}Ba_xMnO_3$ with x = 0.1, 0.15, 0.2 and 0.3. As barium percentage increases, the average A-site ionic radius $\langle r_A \rangle$ and hence the effective cationic mismatch $\sigma^2$ increases [18-20]. This eventually depicts in the shift of XRD peaks (inset of Fig. 1). This can be correlated to the strain induced on $MnO_6$ octahedra with increase in the barium content [21]. Table-I shows that *Pbnm* orthorhombic phase is dominant upto 15% Ba-doping and beyond this the structure gets converted into rhombohedral $R\bar{3}C$ phase. The *hkl* values and Reitveld fitted curves of the observed XRD peaks are shown in Fig. 2 and the calculated lattice parameters have been given in Table-I.

Fig. 3 (a-e) shows the scanning electron micrographs of the pristine and the various Ag-added LCBMO (Ba = 0.1) samples. The presence of Ag has also been confirmed through the elemental analysis (EDAX). A clear grain growth in the samples upto 40% Ag is observed. Average grain size ranges from ~6μm in the pristine sample to a saturation level of ~12μm in the 40% Ag-composite and beyond. Density of the Ag particles (shining white), noticed on the peripheries of the LCBMO grains, has been found to increase with Ag-content. Interestingly in the highest Ag-added sample (80%), segregation of the metallic silver in the form of bunching of crystallites is exhibited (Fig.3e). The increased LCBMO grain growth results in the decrease of the grain boundary density, leading to the enhanced intergrain diffusion [14]. It is worth mentioning here that the differential scanning calorimetric measurements carried out on the 40% Ag-added LCBMO composite has revealed a sharp minimum around 960°C which may be correlated with the melting point of silver. The presence of molten silver during phase formation (like liquid phase sintering) seemingly has acted as catalyst for better grain growth through bulk diffusion process in the composite material.

Fig. 4 exhibits the resistance variation with temperature for $La_{0.7}Ca_{0.3-x}Ba_xMnO_3$ ($0 \leq x \leq 0.20$) series. It is observed that $T_P$ increases from 260K for pure $La_{0.7}Ca_{0.3}MnO_3$ to 306K for $La_{0.7}Ca_{0.1}Ba_{0.2}MnO_3$ sample. It is also worthwhile to notice that though $T_P$ has shifted towards room temperature with Ba-doping, the steepness of the curve, however, has decreased. This manifests from the random tilting of the $MnO_6$ octahedra with increasing Ba-doping at Ca-site which eventually hinders the electron transfer from one Mn-site to the other and weakens the double-exchange mechanism. Therefore, the electrical resistivity of the Ba-doped samples is more than the pristine sample below $T_P$. Inset of Fig. 4 shows the calculated TCR as a function of temperature for x = 0.0, 0.10, 0.15 and 0.20 samples. Maximum TCR of ~ 20% is achieved for the pristine $La_{0.7}Ca_{0.3}MnO_3$ at 250K, which decreases to ~ 3% for $La_{0.7}Ca_{0.2}Ba_{0.1}MnO_3$ and $La_{0.7}Ca_{0.15}Ba_{0.15}MnO_3$ samples. TCR is only 0.52% for $La_{0.7}Ca_{0.1}Ba_{0.2}MnO_3$ sample. The values of $T_P$ and TCR are provided in Table-I.

We now discuss the results of Ag containing LCBMO composites. XRD patterns of Ag-added $La_{0.7}Ca_{0.3-x}Ba_xMnO_3$( x = 0.1, 0.15) series are shown in Fig.5. Besides the characteristic peaks of LCBMO, extra peaks corresponding to metallic silver have also been observed (indicated by *) which have been fitted with $Fm\bar{3}m$ space group of cubic structure. No significant shift in the XRD peaks is observed with silver addition indicating that most of Ag remains at the grain boundaries.

Fig.6 shows the resistivity variation with temperature for the Ag-added $La_{0.7}Ca_{0.2}Ba_{0.1}MnO_3$ samples. The transition temperature $T_P$ remains unchanged with Ag addition. This is contrary to the Ag-added Ti-doped $La_{0.67}Ba_{0.33}MnO_3$ samples reported by Yuan etal. [22] where it was found that $T_P$ shifts at higher temperatures with Ag-addition because of the change in the $Mn^{+3}/Mn^{+4}$ ratio. In the present case, however, it seems that $Mn^{+3}/Mn^{+4}$ ratio does not change, only the microstructural deficiencies seem to get removed. The sharpness of the transition increases tremendously with Ag addition and all curves of Ag-added $La_{0.7}Ca_{0.2}Ba_{0.1}MnO_3$ samples nearly overlap around $T_P$. The electrical resistivity also decreases in the whole temperature range because the presence of silver in the grain boundary (GB) regions would render the GBs more conducting by opening the new conducting channels among the LCBMO grains [14-15].

It is observed from the inset of Fig. 6 that TCR values at 284K remains almost unchanged (~11%) upto 60% of Ag addition. The enhancement of TCR values with Ag addition may be attributed to the following three factors, all of which contribute to the enhancement of the prefactor (1/R) of the TCR formula, viz. grain growth, opening of the new conducting channels and the decrease of the barrier contribution at the grain boundary. Grain boundaries offer a barrier in the carriers flow; therefore, polycrystalline samples exhibit higher resistance than the single crystal. With metal/metal oxide addition the grain boundary density is seen to decrease, resulting in the decrease of the overall contribution of the barrier. Moreover, in the present case Ag addition results in the opening of new conducting paths between the grains (due to high conductivity of silver) which is manifested by the decrease in the overall electrical resistivity.

Though the TCR value of Ag-added $La_{0.7}Ca_{0.2}Ba_{0.1}MnO_3$ samples has increased to 11%, the $TCR^{max}$ temperature (~284K) however, is still lower than the room temperature value. For this reason, we further investigated the properties of the Ag-added $La_{0.7}Ca_{0.15}Ba_{0.15}MnO_3$ series and the results are shown in Fig.7. Like the previous samples, $T_P$ remains invariant (~ 304K) with Ag addition. Inset of Fig. 7 shows increase in TCR values as a function of temperature. The maximum TCR achieved is 5.5% for 80% Ag added sample at 301K. On comparing the TCR values of Ag-added $La_{0.7}Ca_{0.2}Ba_{0.1}MnO_3$ and $La_{0.7}Ca_{0.15}Ba_{0.15}MnO_3$ samples it is noticed that the former composites have high values of TCR though $T_P$ is below room temperature whereas the later have lower TCR values but near to room temperature.

Fig. 8a shows the electrical resistivity variation with temperature of 40% Ag-added $La_{0.7}Ca_{0.2}Ba_{0.1}MnO_3$ sample with and without the application of magnetic field (inset refers to the pristine sample). As expected, the electrical resistivity decreases with magnetic field because the applied field tries to align the localized $t_{2g}$ spins parallel rendering the electron transfer easier. We calculated magneto-resistance (MR) of these and other composites samples using the relation MR% = [(R(0) – R(H)/ R(0)]x100, where R(H) and R(0) are the electrical resistances with and without the application of magnetic field (Fig. 8b). Maximum MR of 19% at 280K is obtained for 40% Ag-added sample as against 11% for the pristine sample. MR at 77K of the pristine sample is above 13%, which goes down to less than 2% for the 80% Ag-added sample.

In polycrystalline manganites total MR is the sum of the contributions from the grains and the grain boundaries. MR near the insulator-metal peak arises from the grains (intragranular or intrinsic) whereas at lower temperatures MR appears due to the grain boundaries (intergranular or extrinsic). Basically, there is spin polarized tunneling of electrons through the barrier at the grain boundary that results in intergranular or extrinsic MR. However, in the composite system the decrease in the grain boundary density will result in the decrease of the spin polarized tunneling and as a consequence of this the corresponding intergranular MR also decreases. The reduction in low temperature MR indicates that the grain boundary effects are reduced with silver addition [14]. The reason for increase of the intrinsic MR with Ag or any other metal/metal oxide addition has been given in terms of the enhanced grain growth and increase in the size and number of the spin clusters [23]. The increase in intrinsic MR with increasing grain size has been reported earlier also [24].

Fig. 8b inset shows the MR variation with magnetic field. Clearly, a linear curve is obtained for both the pristine and $Ag_{0.4}$ samples at 300K and 290K. At a particular field value, MR of 40% Ag-added sample is more than that of pristine sample. This property of linear dependency of MR% on applied field can be used in sensing the magnetic field or as magnetic sensor.

Further, fabrication of thin films of these samples is underway to carry out the noise measurements and to utilize high values of TCR in their thin film form.

## CONCLUSIONS

Enhanced temperature coefficient of resistance (TCR) and magneto-resistance (MR) in Ag-added $La_{0.7}Ca_{0.3-x}Ba_xMnO_3$ manganite composites close to room temperature have been observed. Higher TCR has been attributed to the grain growth and opening of new conducting channels. These factors also contribute in enhancing the intragranular magneto-resistance and simultaneous decrease in intergranular MR. Enhanced TCR and MR at/near room temperature of such materials open up their potential as infra-red bolometric and magnetic field sensors.

## ACKNOWLEDGEMENTS

The authors would like to thank the Director, National Physical Laboratory, New Delhi for his support of the present work, and their colleagues Drs. A. K. Srivastava, H. K. Singh and R. K. Kotnala for their assistance in SEM, Magneto-transport and Magnetization measurements respectively. One of the authors (RT) is thankful to the University Grant Commission (UGC) New Delhi, India or the grant of Senior Research Fellowship. The present work is supported by Indo-German-DST-DAAD-PPP-2008 program. Financial support to the author (SKA) under the Emeritus Scientists Scheme of the Council of Scientific & Industrial Research (CSIR), New Delhi is also gratefully acknowledged.

# FIGURE CAPTIONS

**Fig. 1:** XRD Pattern of $La_{0.7}Ca_{0.3-x}Ba_xMnO_3$ (x = 0.1, 0.15, 0.2, 0.3).

**Fig.2:** Calculated and observed XRD patterns of $La_{0.7}Ca_{0.2}Ba_{0.1}MnO_3$, $La_{0.7}Ca_{0.15}Ba_{0.15}MnO_3$ and $La_{0.7}Ca_{0.2}Ba_{0.1}MnO_3 +Ag_{0.6}$. The Bragg-peak positions are shown for different space groups.

**Figs. 3**(a)-(e)**:** Scanning electron micrographs of $La_{0.7}Ca_{0.2}Ba_{0.1}MnO_3 +Ag_y$ (y = 0.0, 0.2, 0.4, 0.6 and 0.8) samples.

**Fig.4**: Normalized R(T) plots of $La_{0.7}Ca_{0.3}MnO_3$, $La_{0.7}Ca_{0.2}Ba_{0.1}MnO_3$ and $La_{0.7}Ca_{0.15}Ba_{0.15}MnO_3$ and $La_{0.7}Ca_{0.1}Ba_{0.2}MnO_3$ samples. Inset shows the TCR% plots of the same samples.

**Fig. 5:** XRD Pattern of Ag-added $La_{0.7}Ca_{0.2}Ba_{0.1}MnO_3$ and $La_{0.7}Ca_{0.15}Ba_{0.15}MnO_3$ composites

**Fig. 6:** $\rho$ (T) plots of $La_{0.7}Ca_{0.2}Ba_{0.1}MnO_3 +Ag_y$ (y = 0.0, 0.2, 0.4, 0.6 and 0.8) samples. Inset shows the TCR% plots of the samples.

**Fig.7:** $\rho$ (T) plots of $La_{0.7}Ca_{0.15}Ba_{0.15}MnO_3 +Ag_y$ (y = 0.0, 0.2, 0.4, and 0.8) samples; Inset shows the TCR% plots of the samples.

**Fig.8(a) :** $\rho$ (T) for the pristine and 40% Ag-added $La_{0.7}Ca_{0.2}Ba_{0.1}MnO_3$ samples, in the presence and absence of the applied magnetic field.

**Fig.8(b):** MR% as a function of temperature of $La_{0.7}Ca_{0.2}Ba_{0.1}MnO_3$, 40% and 80% Ag-added $La_{0.7}Ca_{0.2}Ba_{0.1}MnO_3$ samples . Inset showing the variation of MR% with applied field for $La_{0.7}Ca_{0.2}Ba_{0.1}MnO_3$ and 40% Ag-added $La_{0.7}Ca_{0.2}Ba_{0.1}MnO_3$ at 290 and 300K.

**Table-I.** Reitveld refined Lattice parameters, $T_P$ and $TCR^{max}$ with corresponding temperature for various samples.

| Compound<br>Refined space group | Lattice Parameter (Å) | $T_P$ (K) | $TCR^{max}$ (%) / T(K) |
|---|---|---|---|
| $La_{0.7}Ca_{0.2}Ba_{0.1}MnO_3$<br>Space group: *Pbnm* | a = 5.5145(2)<br>b = 5.4786(2)<br>c = 7.7391(3) | 289 | 3.2 / 281 |
| $La_{0.7}Ca_{0.15}Ba_{0.15}MnO_3$<br>Space group: $R\bar{3}C$ | a = 5.5145(2)<br>b = 5.5145(2)<br>c = 13.3952(8) | 302 | 3.1 / 299 |
| $La_{0.7}Ca_{0.1}Ba_{0.2}MnO_3$<br>Space group: $R\bar{3}C$ | a = 5.5242(3)<br>b = 5.5242(3)<br>c = 13.4297(9) | 306 | 0.52 / 294 |
| $La_{0.7}Ca_{0.2}Ba_{0.1}MnO_3:Ag_{0.4}$<br>Space group: *Pbnm*/$Fm\bar{3}m$ | a = 5.5149(2)<br>b = 5.4798(3)<br>c = 7.7403(3) | 289 | 11 / 284 |
| $La_{0.7}Ca_{0.2}Ba_{0.1}MnO_3:Ag_{0.8}$<br>Space group: *Pbnm*/$Fm\bar{3}m$ | a= 5.5152(2)<br>b = 5.4800(2)<br>c = 7.7404(3) | 290 | 7.5 / 283 |

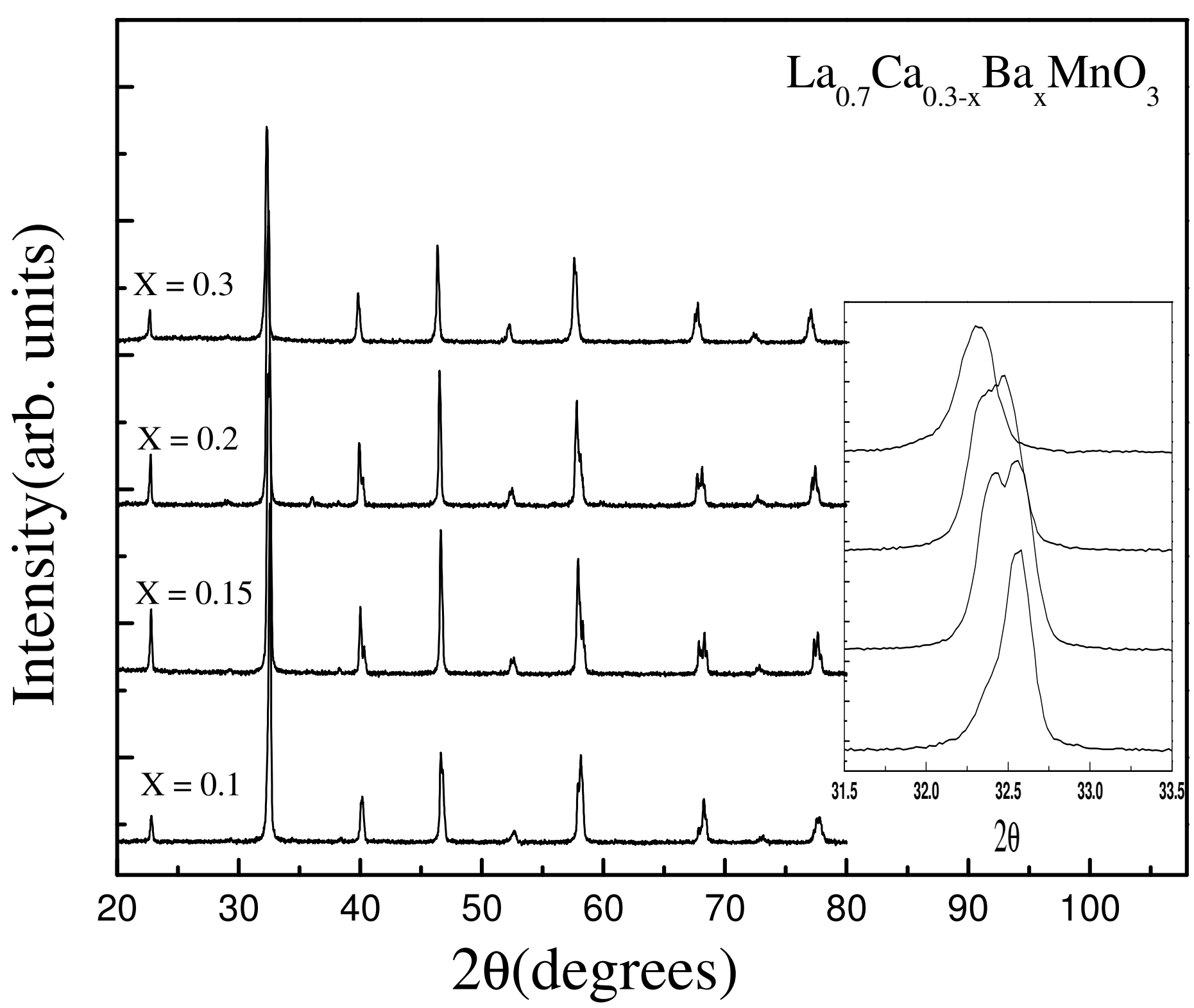

**Figure 1**

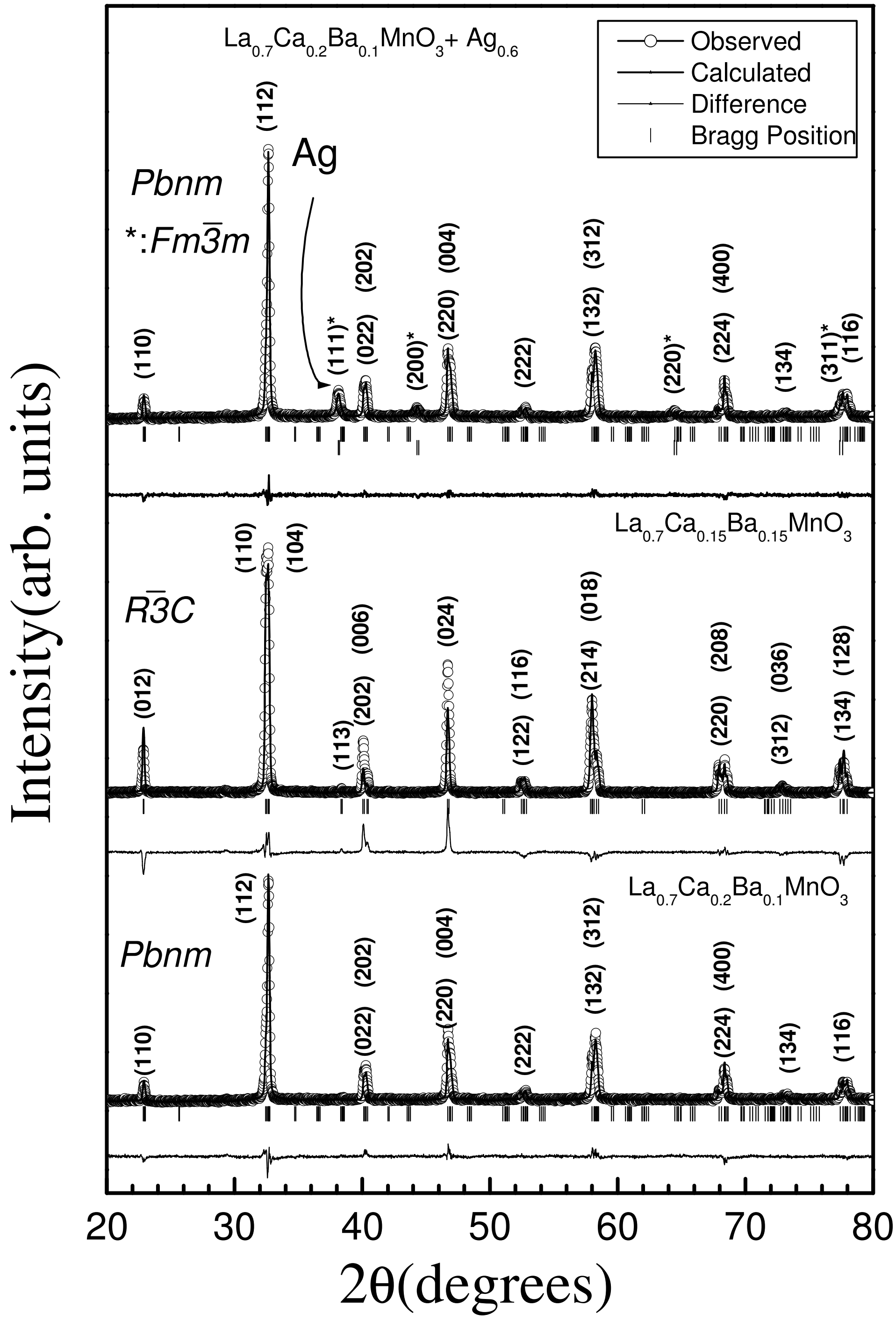

**Figure 2**

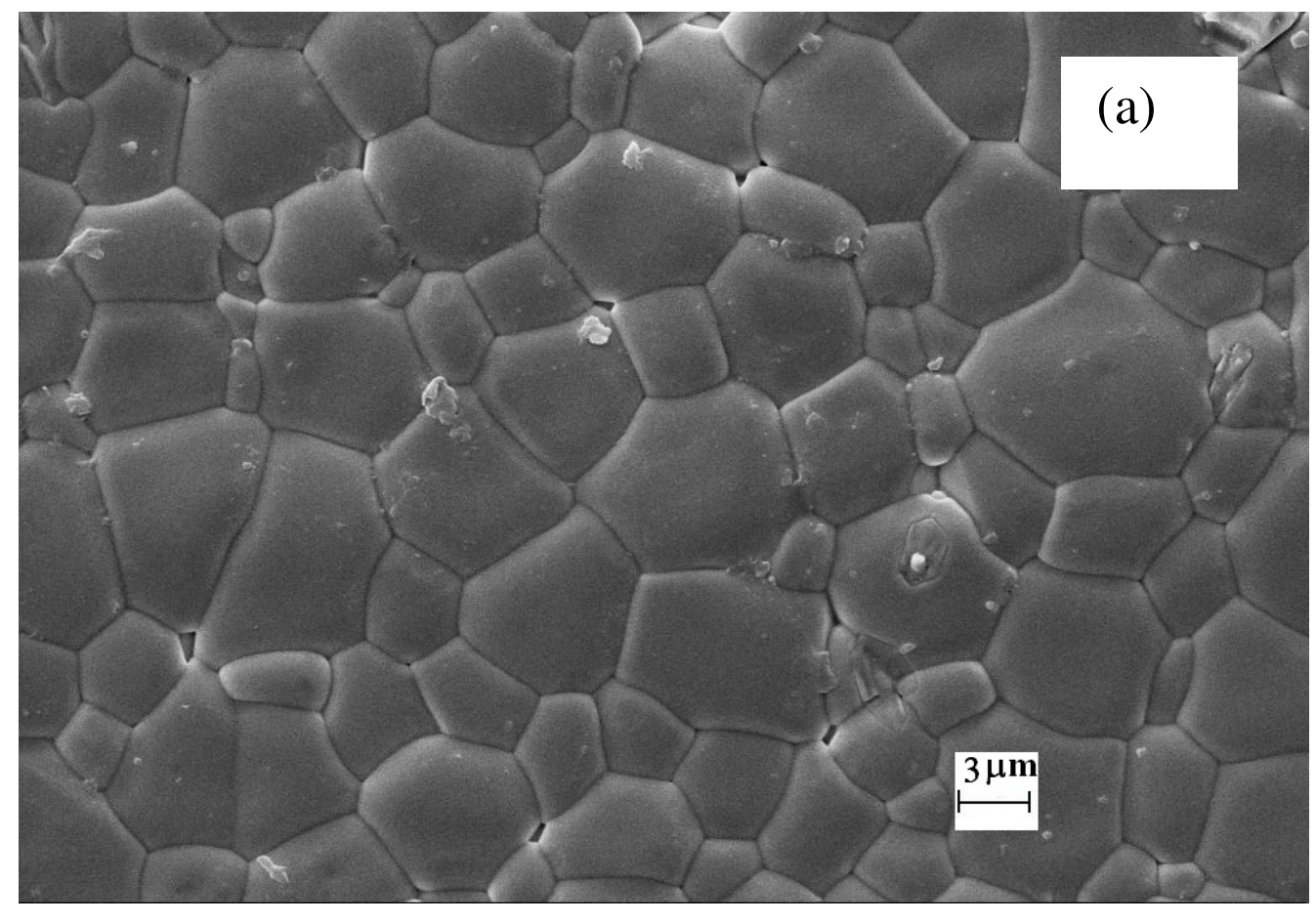

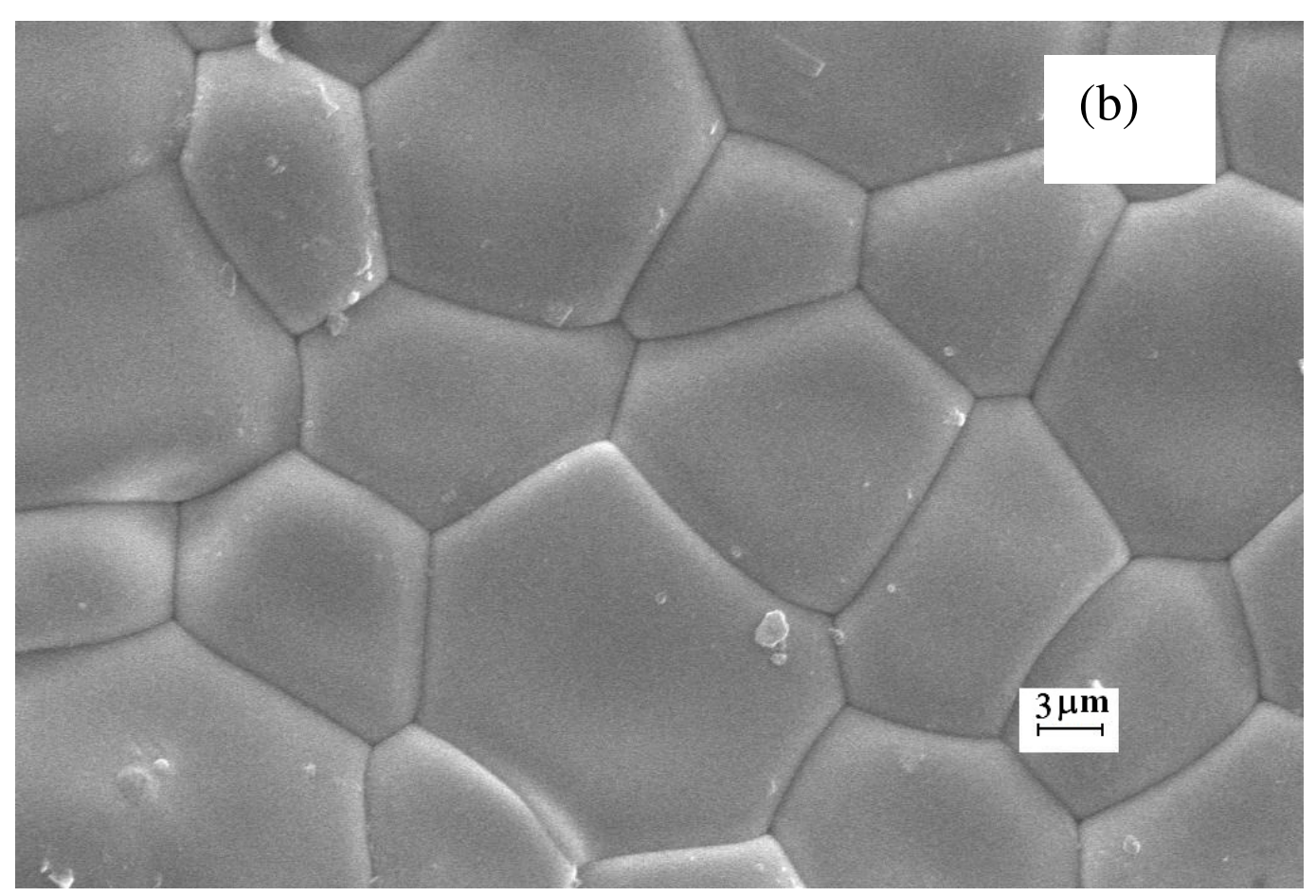

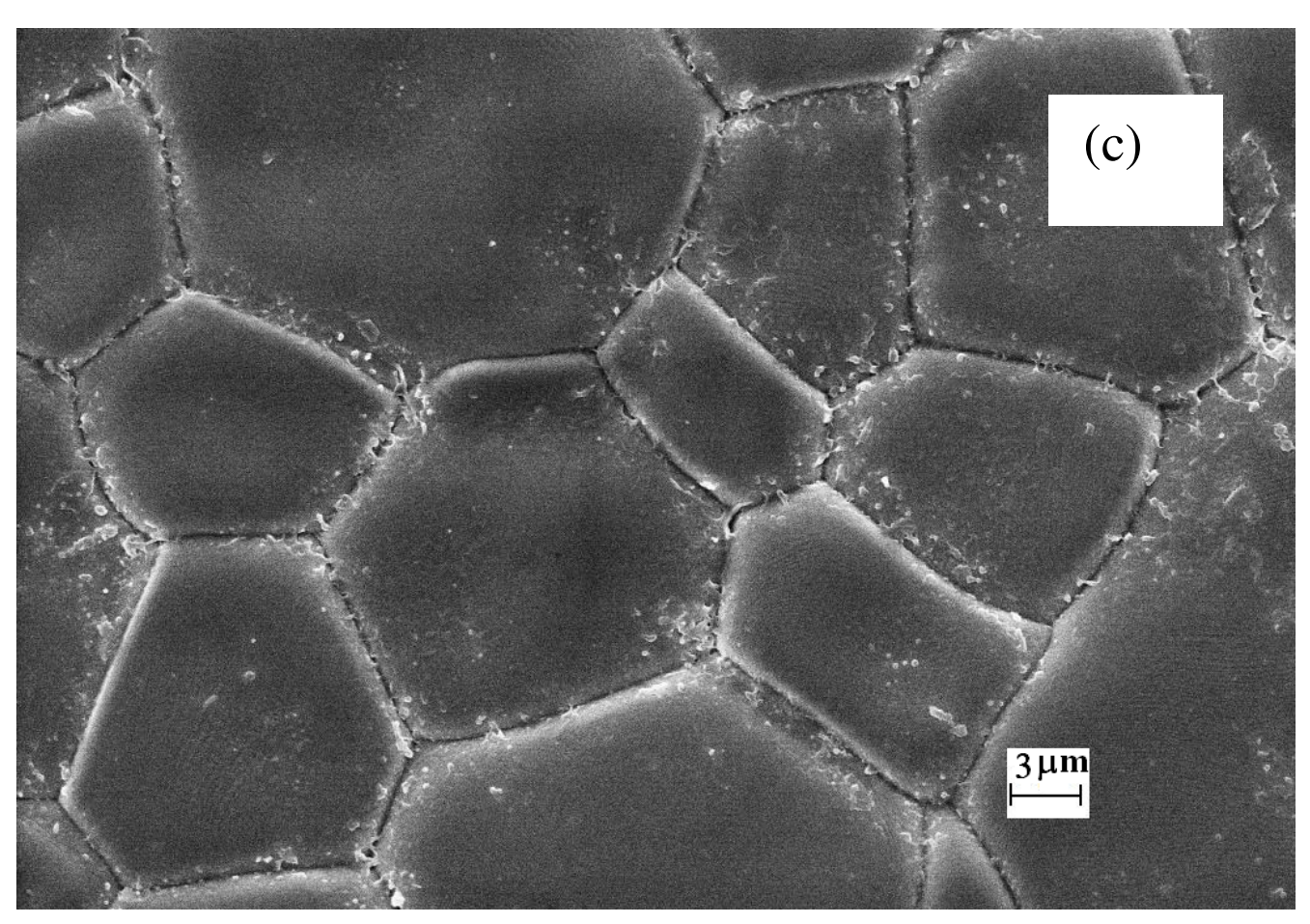

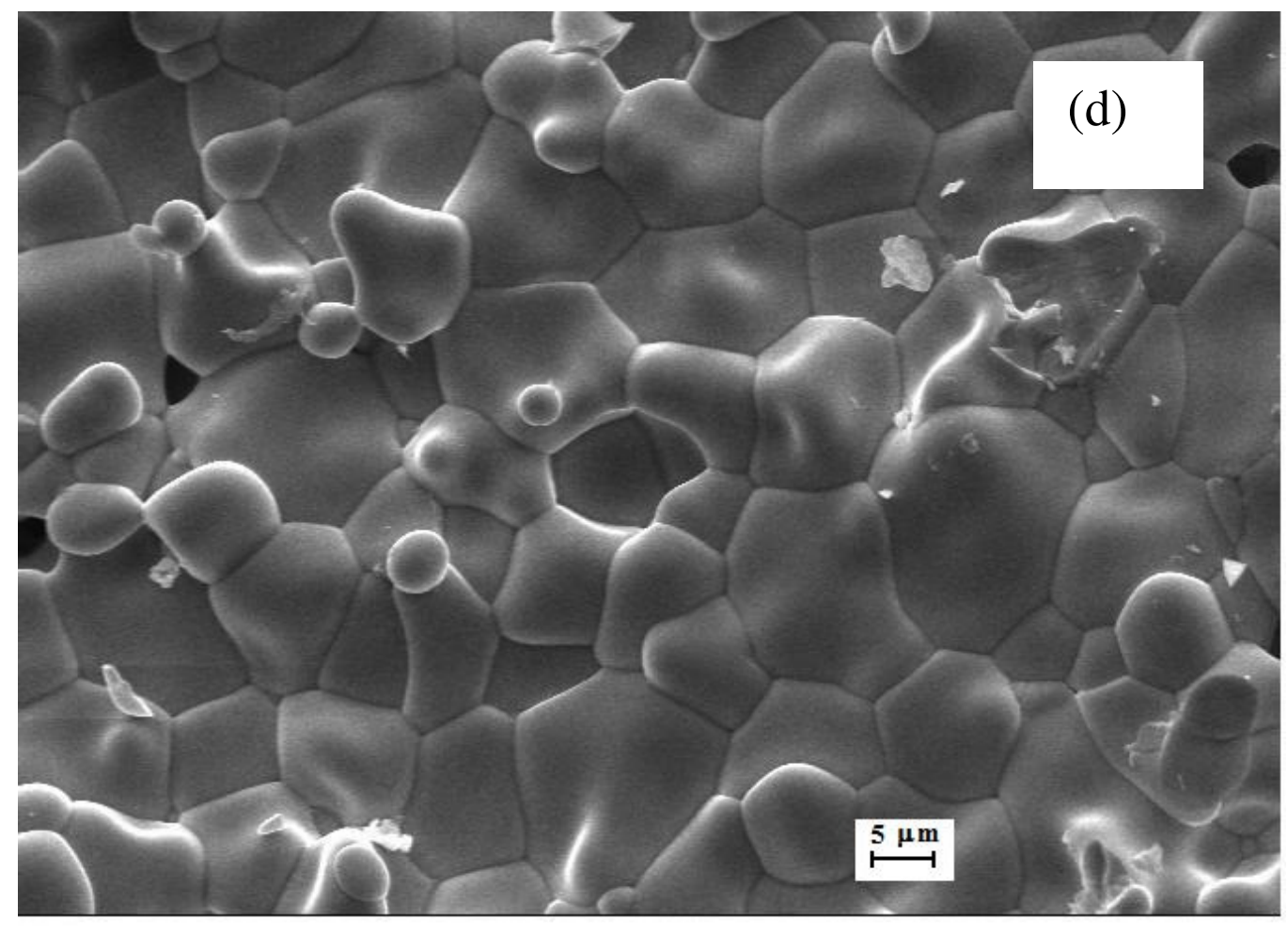

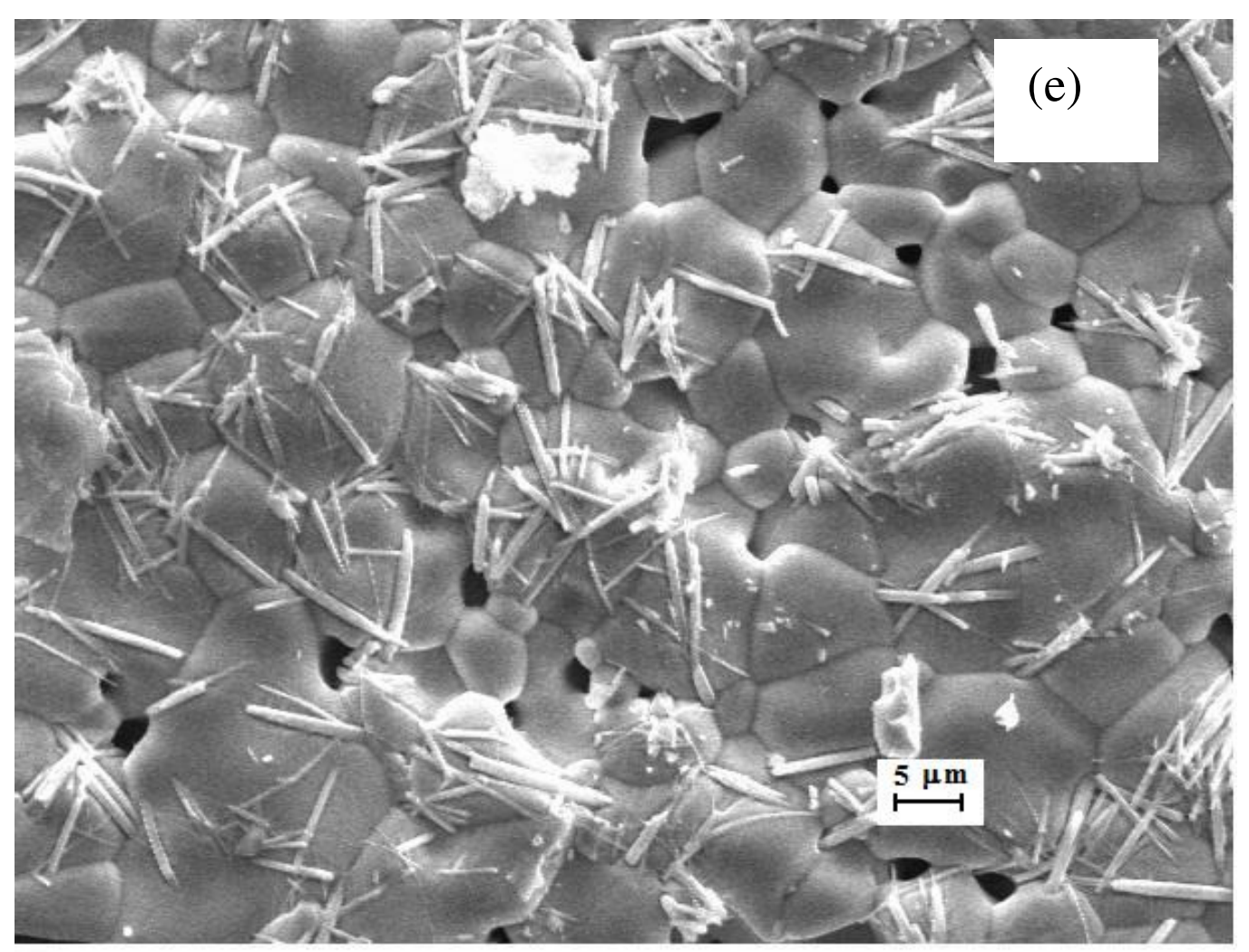

**Figure 3 (a-e)**

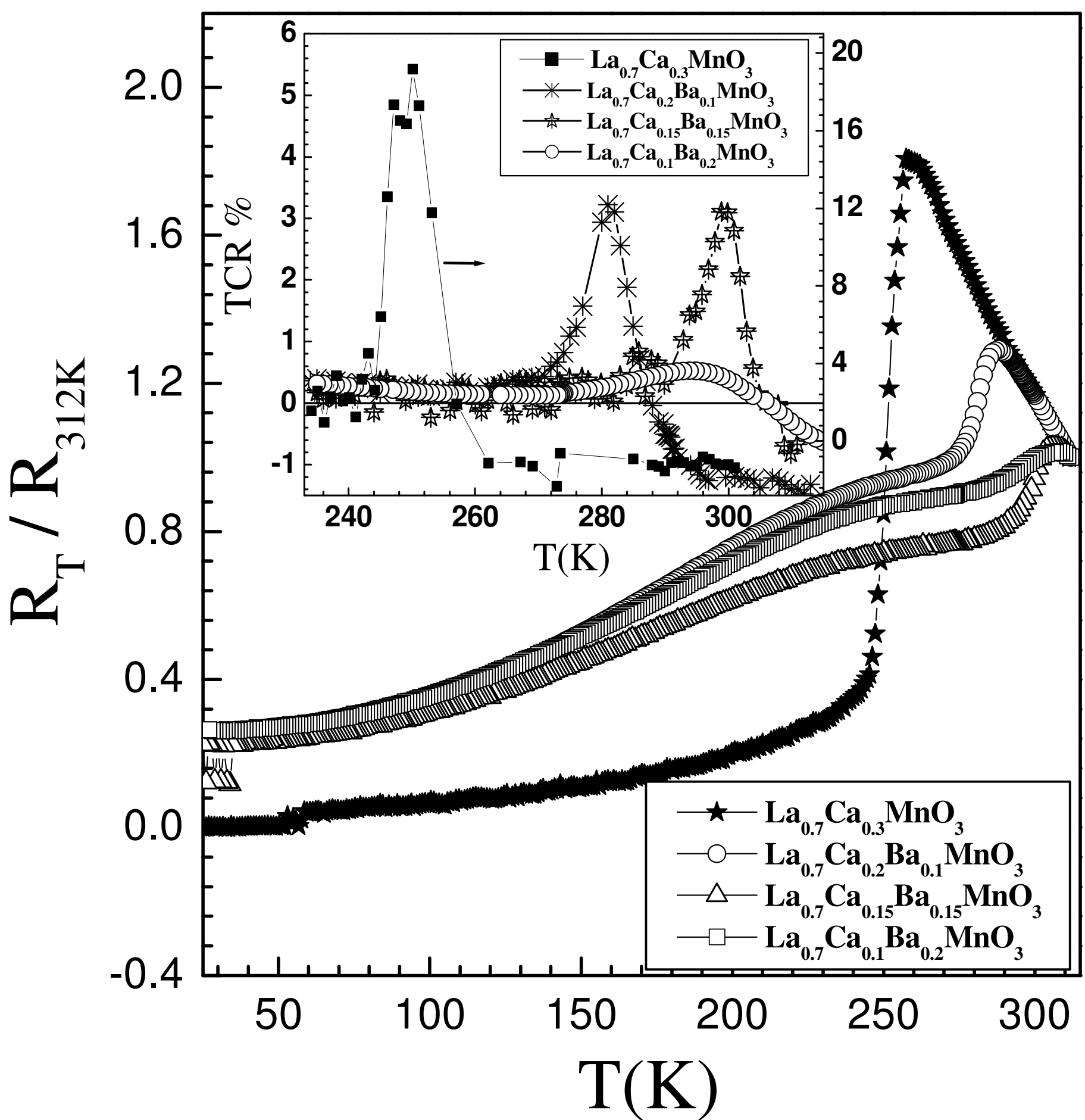

**Figure 4**

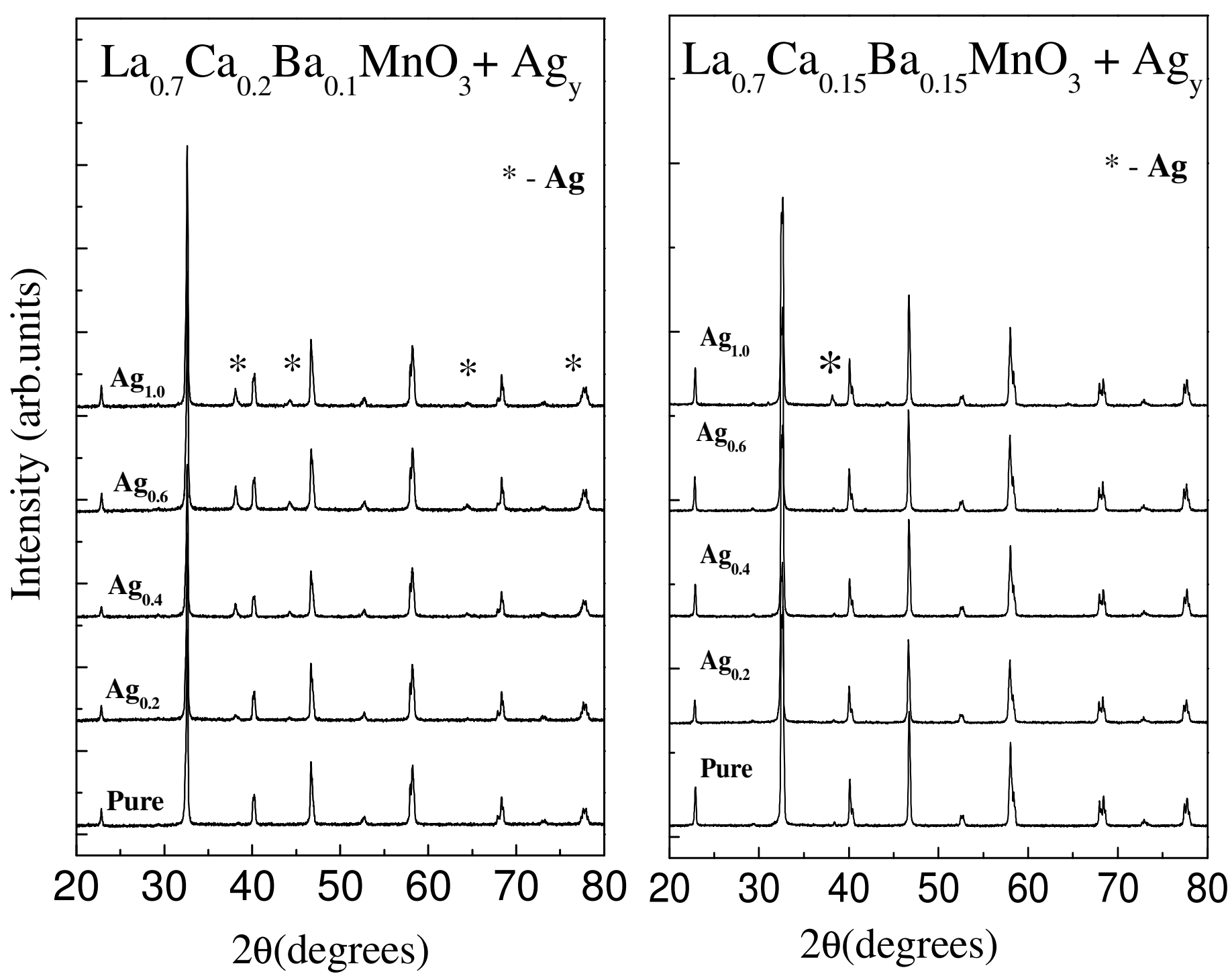

**Figure 5**

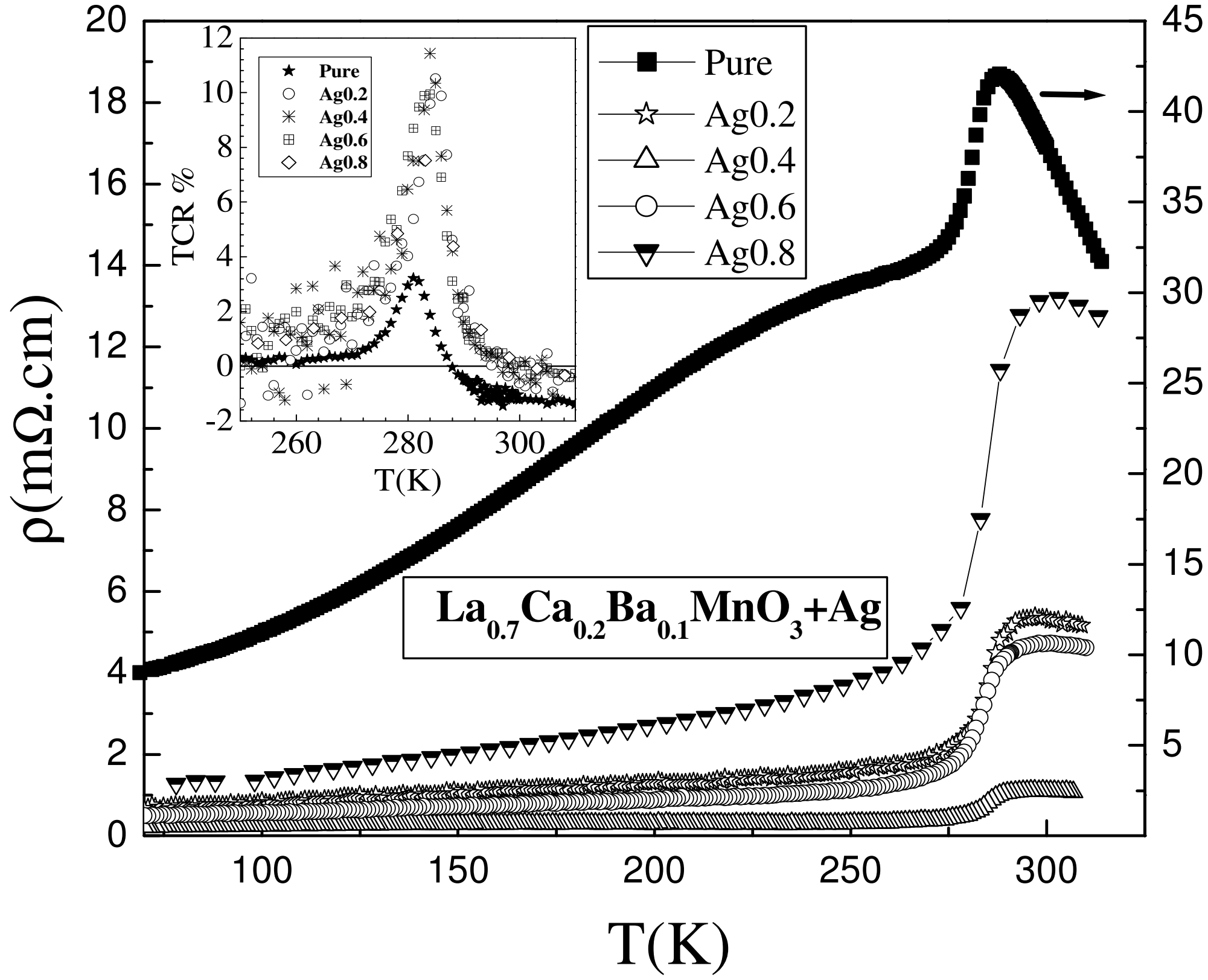

**Figure 6**

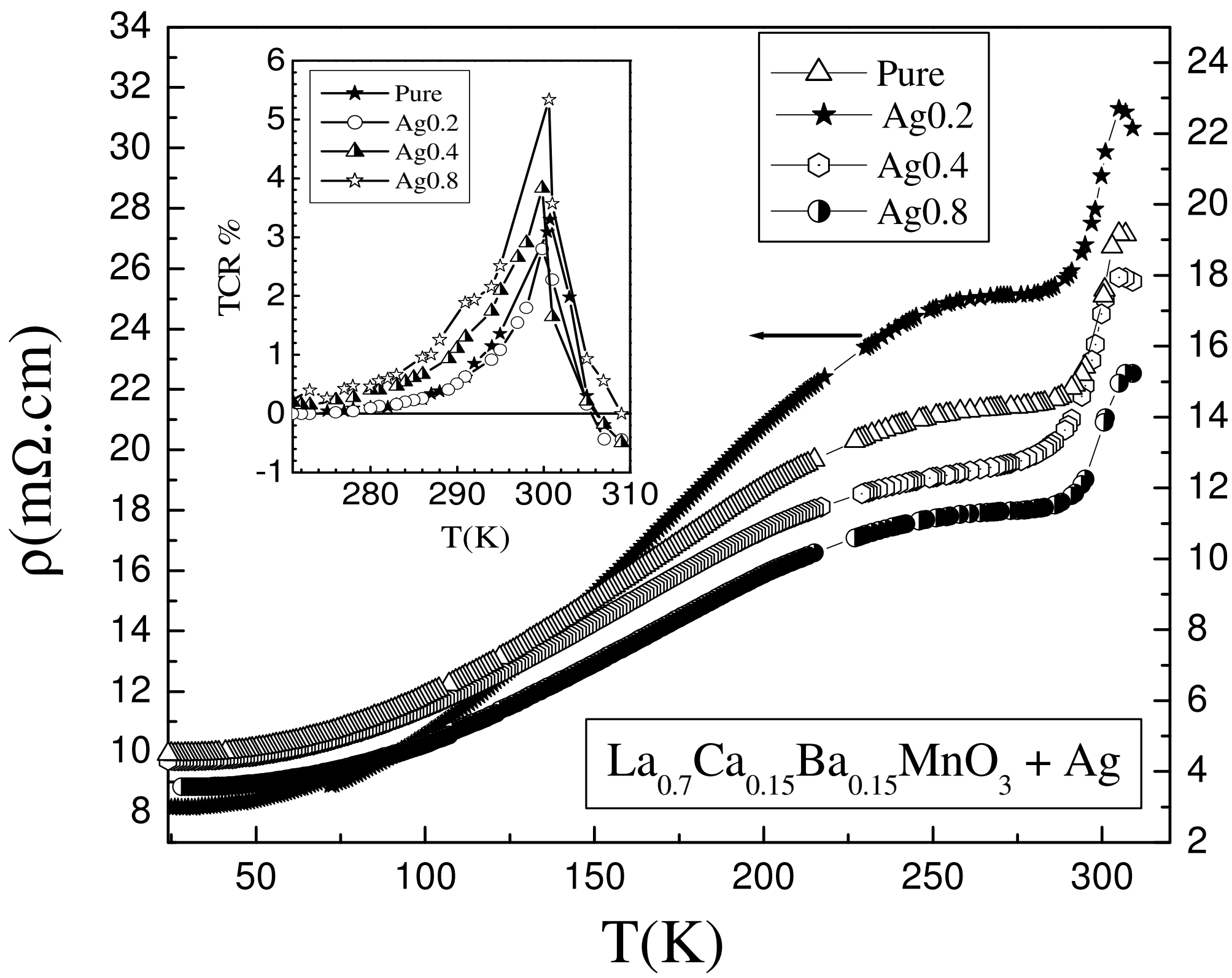

**Figure 7**

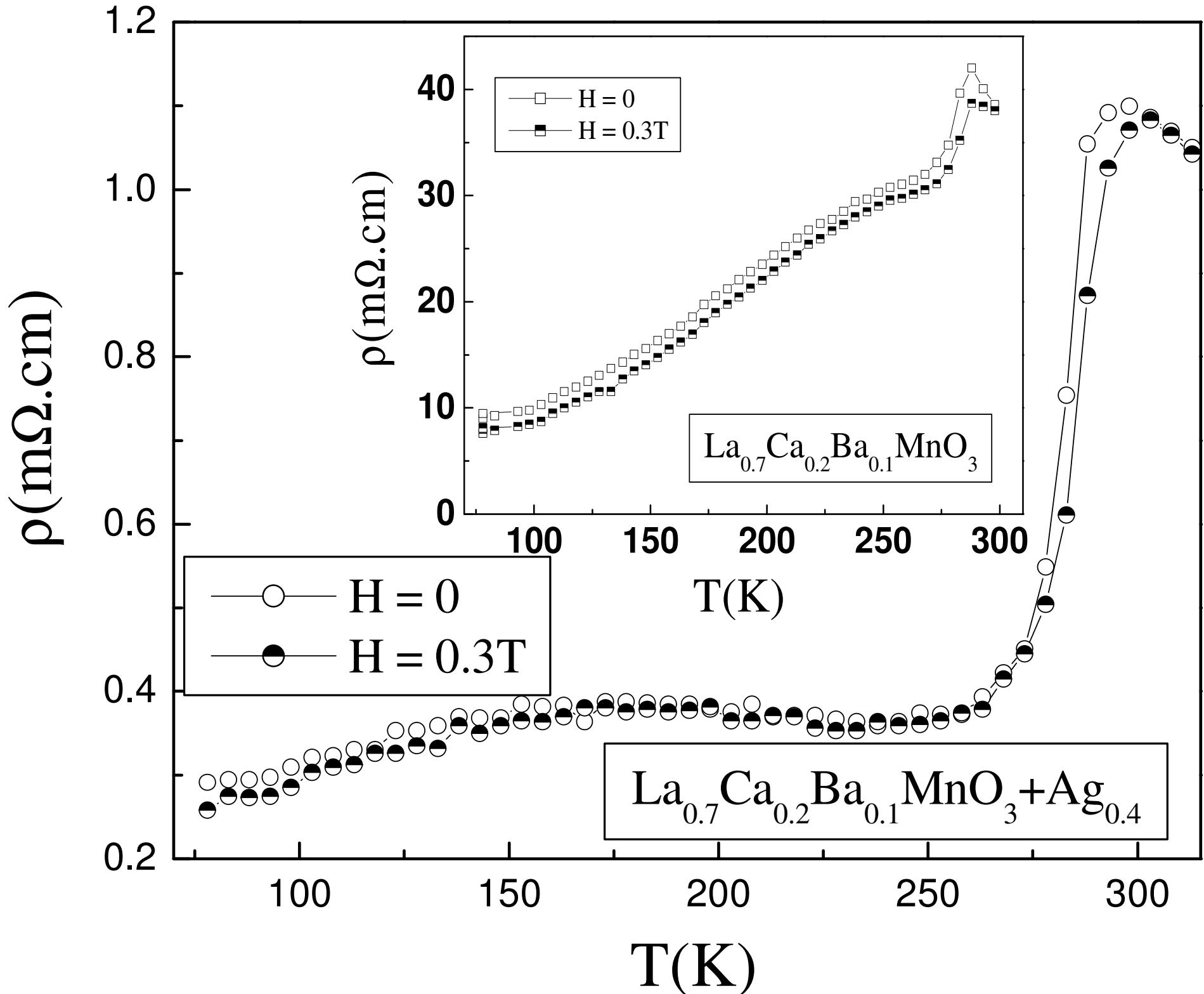

**Figure 8(a)**

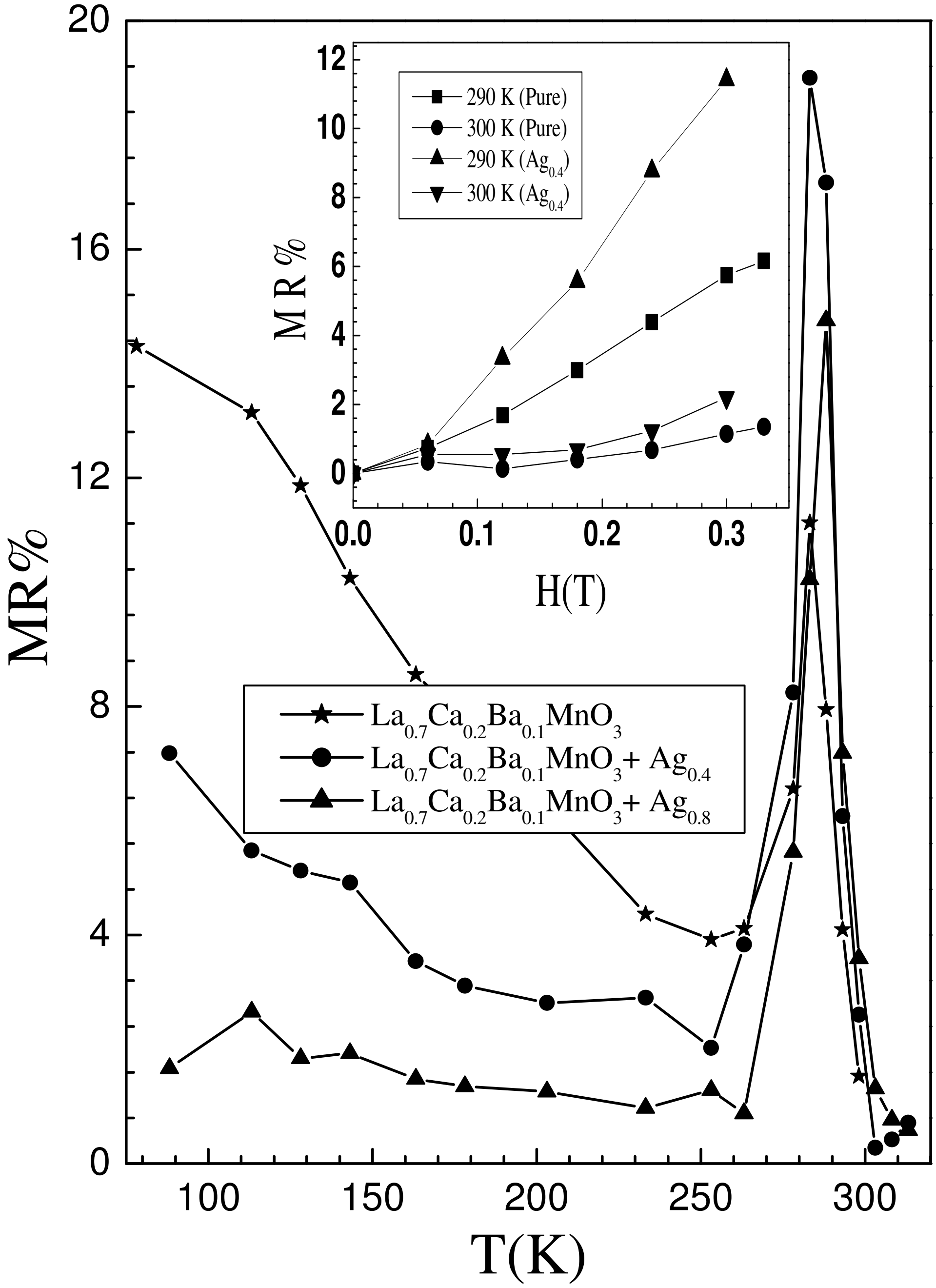

**Figure 8(b)**